\documentclass[%
reprint,
amsmath,amssymb,
aps,longbibliography,
]{revtex4-2}
\usepackage{url}
\usepackage[colorlinks=true, linkcolor=blue,urlcolor=blue,anchorcolor=blue,citecolor=blue,bookmarksnumbered]{hyperref}
\usepackage{graphicx}
\usepackage{dcolumn}
\usepackage{bm}

\usepackage{ulem}   

\begin{document}
	\title{Global adiabatic criterion for fast topological photon transfer in Fock-state lattices}
	
	\author{Jin-Lei~Wu$^{1}$}\email[]{jlwu517@zzu.edu.cn}
	\author{Pei-Yao~Song$^{1}$}	\author{Jia~Li$^{1}$}\author{Ya~Gao$^{1}$}\email[]{ygao@zzu.edu.cn}\author{Yan~Wang$^{2}$}\author{Shi-Lei~Su$^{1,3}$}\email[]{slsu@zzu.edu.cn}
\affiliation{$^{1}$School of Physics, Zhengzhou University, Zhengzhou 450001, China\\
	$^{2}$School of Electronics and Information, Zhengzhou University of Light Industry, Zhengzhou 450001, China\\
	$^{3}$Institute of Quantum Materials and Physics, Henan Academy of Science, Henan 450046, China}


\begin{abstract}
		Topological state transfer in Fock-state lattices has been demonstrated with high speed using sinusoidal profiles of coupling, yet the underlying reason has remained unclear. A global adiabatic criterion (GAC) is developed to bound the infidelity by the mean and variance of the nonadiabatic factor. The GAC reveals that the key to fast transfer is not a constant energy gap but the vanishing nonadiabaticity variance. For power-law coupling profiles, the variance vanishes only for the sinusoidal shape, which is thus globally optimal. Incorporating experimental decoherence parameters, it is predicted that the optimal transfer duration for a five-photon state is 161~ns, far shorter than 600~ns used in the experiment, reducing time by over $73\%$ while increasing transferred photons by $29\%$. The optimal duration follow a simple linear scaling with photon number, providing a practical guideline. Through constructing an alternative constant-gap coupling family, it is confirmed that a constant gap alone is not sufficient for fast topological photon transfer. The essential condition is uniformity of nonadiabaticity. This work offers a rigorous explanation for the observed speed and a general framework for fast topological photonics engineering.
\end{abstract}
\maketitle

\section{Introduction}
Topological photonics, initially inspired by the quantum Hall effect and topological insulators in condensed matter physics, has enabled robust manipulation of classical light through synthetic gauge fields and topological band structures~\cite{LLu2014NatPhoton,Noh2018NatPhoton,Ozawa2019RMP,Ni2023CR,Ehrhardt2023LPR}. Among various phenomena, topological pumping, as a dynamic transport phenomenon of a state across a lattice, has been realized in photonic, acoustic, and cold-atom systems~\cite{Kraus2012PRL,YGKe2016LPR,Lohse2018Nature,Zilberberg2018Nature,Jurgensen2021Nature,YKSun2022NP,Citro2023NRP,Song2024SciAdv,Wu2024NC,HZhang2024LPR,JLWu2026NC,YChen2026LPR}. These achievements rely on classical wave dynamics and do not require field quantization. However, the quantum nature of light, namely its discrete Fock states and bosonic statistics, gives rise to an entirely new class of topological phases with no classical counterpart~\cite{DWWang2016PRL,HCai2020NSR,Saugmann2023PRA,JQLiao2025PRResearch,Naves2025Research}.\medskip

Recent progress in Fock-state lattices~(FSLs), a synthetic dimension platform for topological phases of quantized light, has sparked significant interest~\cite{DWWang2016PRL,HCai2020NSR,JLYuan2021PRB,JLYuan2024AIPX,JFDeng2022Science}. FSLs have been extensively simulated in classical systems such as photonics~\cite{JLYuan2021APLphotonics, CHWu2023PRA,JNZhang2024PRA}, nanomechanics~\cite{TTian2024PRB}, acoustics~\cite{ZGuan2024PRApplied, MPeng2025PRL}, and microwaves~\cite{Yang2024NC}. More uniquely, exotic topological properties have been observed in superconducting quantum circuits~\cite{JFDeng2022Science,JJZhang2025}. In a two-mode Jaynes–Cummings~(JC) model, the FSL reduces to a one-dimensional~(1D) chain that enables topological transfer of photonic Fock states between two cavities by adiabatically varying the coupling ratio. The landmark experiment in a superconducting quantum circuit demonstrated such photon transfer in 600~ns using sinusoidal temporal coupling modulations~\cite{JFDeng2022Science}. The analog implementations of such topological transfer in classical systems~\cite{JLYuan2021APLphotonics, CHWu2023PRA,JNZhang2024PRA,TTian2024PRB,ZGuan2024PRApplied} further proved it to be much faster than the conventional topological pumping~\cite{FMei2018PRA,LQi2020OL,WLiu2022PRA,JNZhang2024PRB,JKGuo2024PRA,JXHan2024PRApplied} in the Su–Schrieffer–Heeger~(SSH) model~\cite{SSH1979PRL}. The acceleration was often attributed to the constant energy gap between the dark state and the bright states. This interpretation, though plausible, obscures the underlying mechanism.\medskip

\begin{figure*}[htp]
	\centering
	\includegraphics[width=0.88\linewidth]{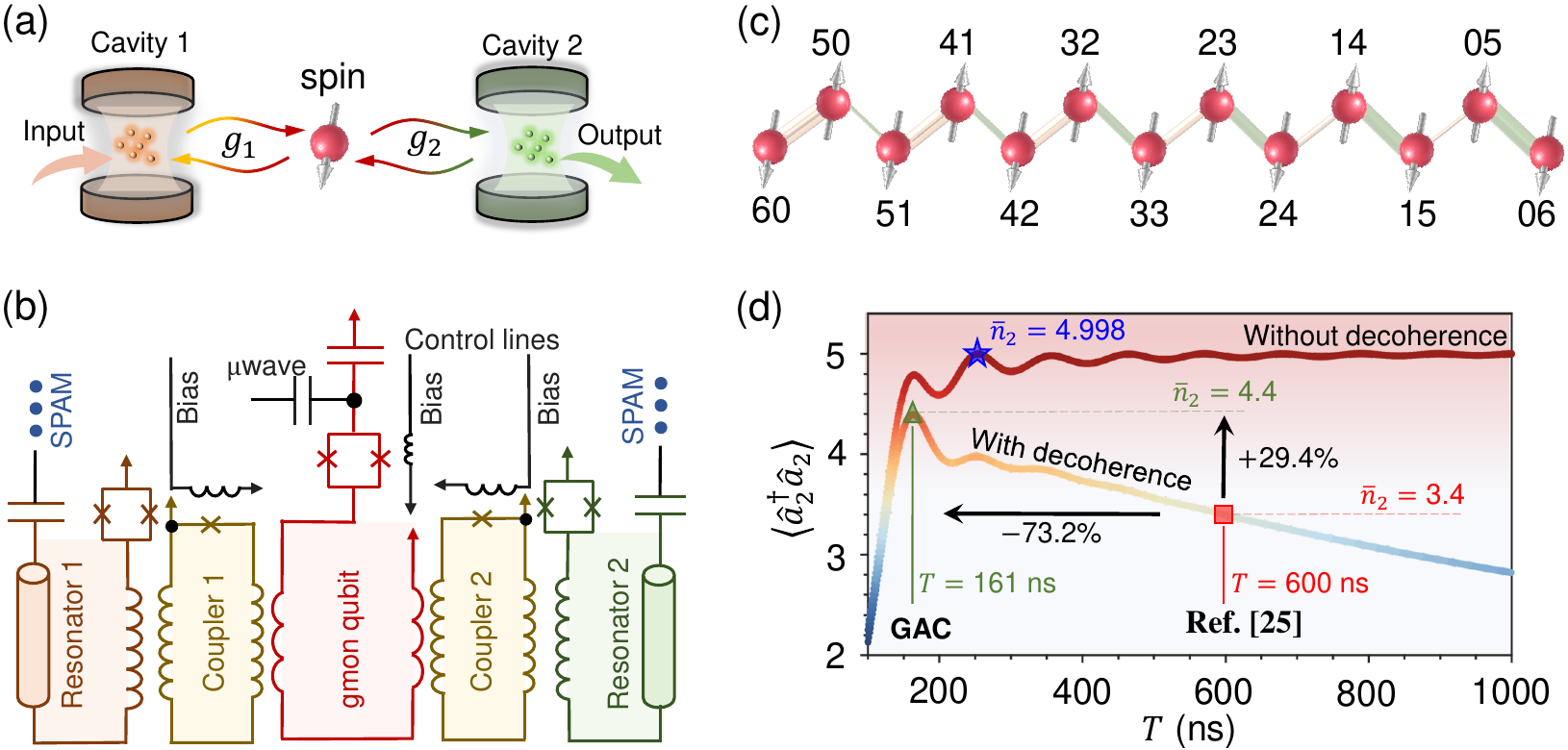}
	\caption{Topological photon transfer in a two-mode JC model. 
		(a) Schematic of the two-mode JC model: two cavity modes (resonators) are coupled to a common spin (qubit) with time-dependent strengths $g_1(t)$ and $g_2(t)$. 
		(b) Superconducting circuit implementation of the device (adapted from {Ref.}~\cite{JFDeng2022Science}), consisting of a central gmon qubit inductively coupled to two resonators via tunable couplers. 
		(c) 1D FSL for total excitation number $N=6$, where each site represents a Fock state $|\downarrow/\uparrow; n_1,n_2\rangle$. The site-dependent hopping amplitudes are proportional to $g_1\sqrt{n_1}$ (brown) and $g_2\sqrt{n_2}$ (green). 
		(d) Final average photon number in the superconducting resonator~2 as a function of duration $T$ for $N=5$ with $\alpha=1$ (sinusoidal profiles of coupling). The blue pentacle marks the optimal duration $T=255$~ns in the absence of decoherence ($\bar n_2=4.998$), while the green triangle indicates the optimal duration $T=161$~ns in the presence of decoherence (peak $\bar n_2=4.5$). The red square shows the experimental point $T=600$~ns~($\bar n_2=3.9$) from {Ref.}~\cite{JFDeng2022Science}.}\label{f1}
\end{figure*}
In this work, we resolve this puzzle by establishing a global adiabatic criterion~(GAC) for FSLs. Instead of requiring slow evolution at every instant, the GAC bounds the total nonadiabatic transition probability by the mean and temporal variance of an instantaneous nonadiabatic factor. This reveals a universal design principle: to minimize infidelity for a given total time, one should shape the time-dependent couplings to make the nonadiabatic factor as uniform as possible, that is, minimizing its fluctuations. We prove that for the power-law profiles of coupling, the variance vanishes only for the first power of sinusoidal shape, which is therefore globally optimal. Building on the GAC, we further incorporate decoherence using actual experimental parameters in {Ref.}~\cite{JFDeng2022Science}, and predict that the optimal transfer duration for a five‑photon state is only 161~ns, far shorter than 600~ns used in the experiment. This optimized duration reduces the transfer time by over $73\%$ while increasing the final transferred photon number by more than $29\%$. Moreover, the optimal durations for the sinusoidal shape follow a simple linear scaling with the total photon number, providing a clear experimental guideline that directly emerges from the competition between unitary pumping dynamics and decoherence effect. In addition, we construct an alternative coupling family that maintains a strictly constant energy gap for any power exponent. It shows that non‑optimal exponents lead to larger infidelity due to non‑zero variance, and confirms that uniformity of the nonadiabatic factor, instead of the constant gap, is essential. Our GAC thus offers the first quantitative explanation for the acceleration observed in a series of previous experiments~\cite{JFDeng2022Science,CHWu2023PRA,ZGuan2024PRApplied,TTian2024PRB}, and could establish a general framework for fast, high-fidelity topological photonics engineering in FSLs and further in quantum information processing~\cite{PYSong2025arxiv}.

\section{FSL in a two-mode JC model}
The topological photon transfer in 1D FSL is implemented based on a two-mode JC model. As shown in Fig.~\ref{f1}a, the system contains two cavity modes (labeled 1 and 2) coupled to a common spin, with temporally modulated single-photon coupling strengths $g_1(t)$ and $g_2(t)$, respectively. The landmark experiment by Deng et al. realized this model for 1D FSL topological pumping in a superconducting circuit~\cite{JFDeng2022Science}. The device, as schematically shown in Fig.~\ref{f1}b, consists of a central gmon qubit~\cite{Roushan2017NP} (acting as the spin) inductively coupled to two resonators (cavities 1 and 2) via tunable couplers, along with ancilla qubits for state preparation and measurement~(SPAM). The central qubit and the resonators are all frequency-tunable, allowing precise control over the system parameters. Under the rotating-wave approximation and with all modes resonant with the spin, the Hamiltonian in the interaction picture is~(assuming $\hbar\equiv1$)
\begin{equation}
\hat H(t) = \sum_{j=1}^{2}g_j(t)\,{\hat\sigma}^+\hat a_j + \text{H.c.}
\end{equation}
The raising and lowering operators of the spin are defined by ${\hat\sigma}^+ = |\uparrow\rangle\langle\downarrow|$ and ${\hat\sigma}^- = |\downarrow\rangle\langle\uparrow|$ respectively. $\hat a_j$ denotes the annihilation operator of the cavity $j$. The total excitation number $\hat N = \hat a_1^\dagger\hat a_1 + \hat a_2^\dagger\hat a_2 + |\uparrow\rangle\langle\uparrow|$
is conserved. For a fixed total excitation number $N$, the product Fock states $|\downarrow; n_1,n_2\rangle$ (with $n_1+n_2 = N$), $|\uparrow; n_1-1,n_2\rangle$ and $|\uparrow; n_1,n_2-1\rangle$ form a 1D chain, i.e., FSL as shown in Fig.~\ref{f1}c. The hopping amplitudes are site-dependent
\begin{eqnarray}
\langle\uparrow; n_1-1,n_2|\hat H(t)|\downarrow; n_1,n_2\rangle = g_1(t)\sqrt{n_1},\nonumber\\
\langle\uparrow; n_1,n_2-1|\hat H(t)|\downarrow; n_1,n_2\rangle = g_2(t)\sqrt{n_2}.
\end{eqnarray}
These couplings reflect the bosonic enhancement factor $\sqrt{n}$, that is, the more photons in a cavity, the stronger its effective coupling to the spin.\medskip

Introduce the instantaneous bright and dark modes of the two cavities
\begin{eqnarray}
\hat b_0(t) = \sin\theta(t)\,\hat a_1 + \cos\theta(t)\,\hat a_2,\nonumber\\
\hat b_1(t) = \cos\theta(t)\,\hat a_1 - \sin\theta(t)\,\hat a_2,
\end{eqnarray}
with the total single-photon coupling strength $G(t) \equiv \sqrt{g_1(t)^2+g_2(t)^2}$ and the mixing angle $\theta(t)=\arctan[g_1(t)/g_2(t)]$.
Then the Hamiltonian becomes
\begin{equation}\label{Hb}
\hat H_b(t) = G(t)\,\bigl({\hat\sigma}^+\hat b_0 + {\hat\sigma}^-\hat b_0^\dagger\bigr),
\end{equation}
decoupled to the dark mode $\hat b_1$. For a fixed total excitation number $N$, the dark state (zero‑energy eigenstate) is given by  $|\psi_{\text{dark}}(t)\rangle = |\downarrow; 0, N\rangle_b$.
The subscript $b$ denotes the Fock basis of $(\hat b_0,\hat b_1)$, where the first entry is the photon number in $\hat b_0$, the second in $\hat b_1$. In the local cavity basis $\{|\downarrow; n_1,n_2\rangle\}$, this state expands into a binomial superposition
\begin{eqnarray}
|\psi_{\text{dark}}(t)\rangle &=& \sum_{n=0}^{N} \sqrt{\frac{N!}{n!(N-n)!}}\,
\cos^{n}\theta(t)[-\sin\theta(t)]^{N-n}\nonumber\\
&&\times|\downarrow; n, N-n\rangle\\
\xrightarrow{\text{mapping}} &&\sum_{n=0}^{N} p_{2n+1}(t) |2n+1\rangle_{\text{FSL}}.\nonumber \label{eq:dark_binomial}
\end{eqnarray}
where $p_{2n+1}(t)$ is the instantaneous probability amplitude of the $(2n+1)$-th site $|2n+1\rangle_{\text{FSL}}=|\downarrow; n, N-n\rangle$ in the 1D FSL. This representation makes explicit the coherent distribution of photons between the two cavities and directly connects the dark state to a wave packet in the FSL.\medskip

The topological pumping of photons from cavity 1 to cavity 2 is realized by adiabatically varying the mixing angle $\theta(t)$ from $0$ (where $g_2 \gg g_1$) to $\pi/2$ (where $g_2 \ll g_1$) over a total duration $T$. Throughout the evolution the system stays in the dark state, hardly populating the bright states with the excited spin, because the energy gap $\Delta E(t) = G(t)$ protects the adiabaticity. The conventional instantaneous adiabatic condition requires $|\dot\theta(t)| \ll \Delta E(t)$ pointwise. We will show in the following that a global criterion of adiabatic evolution accounts for fast topological transfer by exploiting the uniformity of $\sqrt N|\dot\theta(t)|/\Delta E(t)$.

\section{Topological photon transfer with global adiabatic criterion}
\label{sec:GAC}
To incorporate our GAC theory to topological pumping in FSLs, we consider a coupling family
\begin{equation}\label{coupling}
g_1(t) = g_0\left|\sin^\alpha\left(\frac{\pi t}{2T}\right)\right|,\quad
g_2(t) = g_0\left|\cos^\alpha\left(\frac{\pi t}{2T}\right)\right|,
\end{equation}
where $g_0$ is the maximum coupling strength and the power exponent $\alpha>0$ is a design parameter to be determined. Conventional topological pumping schemes based on the SSH model typically adopt $\alpha=2$, which keeps slow adiabatic evolution with constant $g_1(t)+g_2(t)$~\cite{FMei2018PRA,LQi2020OL,WLiu2022PRA,JNZhang2024PRB,JKGuo2024PRA,JXHan2024PRApplied}. In contrast, topological pumping in FSLs employs $\alpha=1$ to satisfy $g_1(t)^2+g_2(t)^2 = \text{constant}$, a condition that leads to time-independent eigenenergies, as demonstrated in {Refs.}~\cite{JFDeng2022Science,JLYuan2021APLphotonics,ZGuan2024PRApplied,TTian2024PRB,CHWu2023PRA}. With $\alpha=1$, the pumping process is claimed to be fast~\cite{JFDeng2022Science,JLYuan2021APLphotonics,ZGuan2024PRApplied,TTian2024PRB,CHWu2023PRA}. We identify that a five-photon Fock state can be fully transferred from cavity~1 to cavity~2 with duration $T=255$~ns~(blue pentacle, $\bar n_2=4.998$) in the absence of decoherence, as shown in Fig.~\ref{f1}d using the experiment parameter $g_0/2\pi=9$~MHz~\cite{JFDeng2022Science}. This acceleration is often attributed to the constant energy gap between the dark state and the bright states, as stated in {Refs.}~\cite{JFDeng2022Science,JLYuan2021APLphotonics,ZGuan2024PRApplied,TTian2024PRB,CHWu2023PRA}. Contrary to this common belief, in this work our GAC reveals that the essential reason for the acceleration is the uniformity of a defined nonadiabaticity factor, rather than the constancy of the energy gap.\medskip

\begin{figure}[htp]
	\centering
	\includegraphics[width=\linewidth]{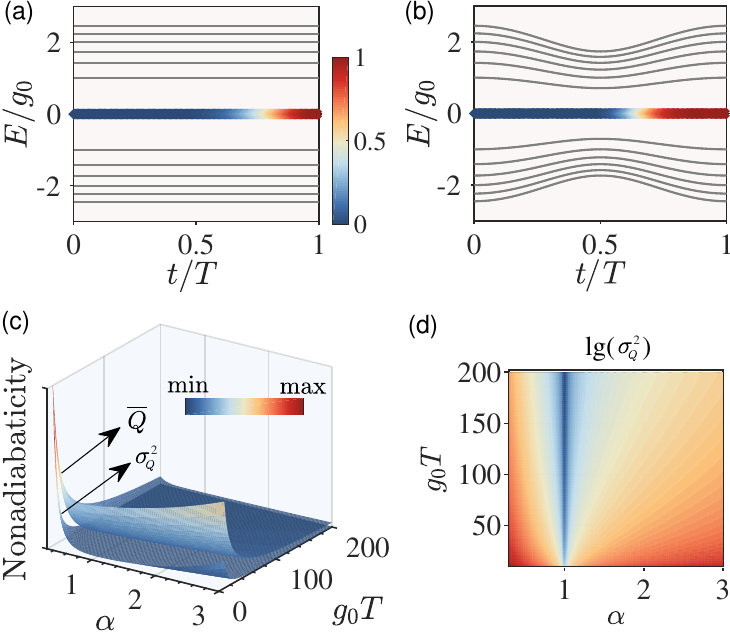}
	\caption{Global adiabatic criterion applied to the coupling family in {Equation}~\eqref{coupling}. 
		(a) Time-independent energy spectrum for $\alpha=1$ (constant gap $\Delta E=\sqrt{N}g_0$) for a closed six-excitation subspace (13-site FSL). Color on the zero-energy line indicates $|\langle \downarrow;0,6|\psi_{\text{dark}}(t)\rangle|^2$.
		(b) Spectrum for $\alpha=2$, showing an apparent gap closing tendency at $t=T/2$ where the nonadiabatic coupling peaks. 
		(c) Mean nonadiabaticity factor $\overline{Q}$ (color scale) as a function of $\alpha$ and dimensionless evolution duration $g_0T$. Both $\overline{Q}$ and $\sigma_Q^2$ decrease with increasing $g_0T$. 
		(d) Logarithmic plot of the variance $\sigma_Q^2$ in the $\alpha$-$g_0T$ plane, confirming that $\sigma_Q^2=0$ exactly at $\alpha=1$. For $\alpha\neq1$, $\sigma_Q^2>0$, indicating larger nonadiabatic fluctuations.}\label{f2}
\end{figure}
The nonadiabatic coupling between the dark state $|\psi_{\text{dark}}\rangle$ and the bright state $|\psi_{\text{bright}}\rangle$ is given by\\ $\langle\psi_{\text{bright}}|\partial_t\psi_{\text{dark}}\rangle =\sqrt{N/2} \dot\theta(t)$, and the energy gap is $\Delta E(t) = G(t)$. According to our GAC theory~\cite{xiao2026acceleratedtopologicalpumpingphotonic}, the nonadiabaticity factor for this two-mode JC model is
\begin{equation}\label{Qnon}
Q(t) = \frac{|\langle\psi_{\text{bright}}|\partial_t\psi_{\text{dark}}\rangle|}{\Delta E(t)} = \frac{\sqrt{N}|\dot\theta(t)|}{\sqrt2G(t)}.
\end{equation}
It indicates that the nonadiabaticity factor $Q(t)$ is directly proportional to $\sqrt{N}$. This reveals that a larger photon number fundamentally intensifies the nonadiabatic transitions within the system, thereby necessitating a correspondingly longer evolution duration to preserve adiabaticity. This theoretical deduction establishes an intrinsic scaling behavior that will be explicitly verified in our subsequent analysis of decoherence.	
The GAC provides a sufficient, albeit conservative, condition for the infidelity $\varepsilon$ to stay below a threshold $\varepsilon_c$~\cite{xiao2026acceleratedtopologicalpumpingphotonic}
\begin{equation}\label{GAC}
\Delta E_{\min}\,T\,\sqrt{\overline{Q}^2 + \sigma_Q^2} \le \sqrt{\varepsilon_c},
\end{equation}
where $\overline{Q} \equiv \frac{1}{T}\int_0^T Q(t)\mathrm{d}t$ and $\sigma_Q^2 \equiv \frac{1}{T}\int_0^T [Q(t)-\overline{Q}]^2 \mathrm{d}t$ are the mean and variance of $Q(t)$ over $t\in[0,T]$. $\Delta E_{\min}$ is the minimum energy gap. This shows that the upper bound on infidelity scales as $(\Delta E_{\min}T)^2(\overline{Q}^2+\sigma_Q^2)$. Naturally, it is desirable to reduce the fluctuation $\sigma_Q^2$ so as to allow a large mean nonadiabaticity factor (hence a short total time $T$) while maintaining high fidelity. Physically, the temporal variance $\sigma_Q^2$ characterizes the non-uniformity or abrupt fluctuations of the non-adiabatic driving throughout the evolution. From the perspective of time-dependent perturbation theory, a non-zero variance implies the presence of localized sharp peaks or rapid variations in the non-adiabatic factor $Q(t)$. Such violent temporal fluctuations introduce high-frequency components that can more effectively bridge the instantaneous energy gap, thereby severely stimulating unwanted transitions from the dark state to the bright states. By minimizing $\sigma_Q^2$ to zero, the non-adiabatic driving is distributed completely uniformly over the entire duration $T$. This suppresses localized non-adiabatic ``hot-spots'' and prevents severe transition risks, fundamentally explaining why a uniform coupling profile tightly bounds the infidelity and enables an accelerated state transfer. Consequently, for any given coupling family, minimizing the temporal fluctuations of $Q(t)$ should be pursued as a primary design goal.\medskip

The total change of $\theta$ from $t=0$ to $t=T$ is $\Delta\theta = \pi/2$, hence $\int_0^T |\dot\theta|\mathrm{d}t = \pi/2$. The mean value of $Q$ is $\overline{Q} = \frac{1}{T}\int_0^T {{\sqrt{N/2}|\dot\theta(t)|}/G(t)}\mathrm{d}t$. 
For the coupling family under consideration, $\Delta E(t)=G(t)=g_0\sqrt{\sin^{2\alpha}\phi+\cos^{2\alpha}\phi}$ with $\phi=\pi t/(2T)$. The crucial observation is that the variance $\sigma_Q^2$ strongly depends on $\alpha$. When $\alpha=1$, $Q(t) = \sqrt{N}{\pi}/{2\sqrt2g_0T}$ is a constant, yielding $\sigma_Q^2=0$, with a time-independent spectrum as shown in Fig.~\ref{f2}a for the system confined in a closed six-excitation Hilbert subspace~(a 13-site FSL). For any $\alpha\neq1$, $Q(t)$ becomes non‑constant. It develops a peak at $\phi=\pi/4$ when $\alpha>1$ (since the factor $\sin^{\alpha-1}\phi\cos^{\alpha-1}\phi$ peaks at the center) or a dip when $\alpha<1$, resulting in $\sigma_Q^2 > 0$. As an illustrative example, we show the spectrum in Fig.~\ref{f2}b for the case of $N=6$ and $\alpha=2$. It shows an apparent gap closing tendency at $\phi=\pi/4$.
Overall, according to the GAC bound, for a given total evolution duration $T$, the infidelity $\varepsilon$ is minimized (or equivalently, for a given $\varepsilon$ the required $T$ is minimized) when $\alpha=1$. This provides the theoretical justification for the optimality of sinusoidal shapes, and exactly explains why the coupling profiles used in {Refs.}~\cite{JFDeng2022Science,JLYuan2021APLphotonics,ZGuan2024PRApplied,TTian2024PRB,CHWu2023PRA} implement fast topological pumping.\medskip

The optimal $\alpha$ that yields the smallest variance provides the most uniform nonadiabaticity factor, which is the design goal advocated by our GAC theory. By varying the dimensionless parameters $\alpha$ and the total evolution time $T$ (scaled by $g_0$), we show in Fig.~\ref{f2}c the numerical results for the mean $\overline{Q}$ and the variance $\sigma_Q^2$ of the nonadiabaticity factor. In addition to the expected behavior that both $\overline{Q}$ and $\sigma_Q^2$ decrease with increasing $g_0 T$ (i.e., longer evolution or stronger coupling), it is identified that there is an optimal value of $\alpha$ that minimizes the nonadiabaticity in the topological pump. For minimizing the mean $\overline{Q}$, the optimal value is around $\alpha=1$, which is also the point where the energy gap is constant. More importantly, for minimizing the variance $\sigma_Q^2$, as shown in the logarithmic plot of $\sigma_Q^2$ in Fig.~\ref{f2}d, we determine that the optimal value is exactly $\alpha=1$, where $\sigma_Q^2=0$ identically. For $\alpha\neq1$, the variance becomes positive, indicating larger fluctuations of $Q(t)$ and thus a larger infidelity bound. Hence, a faster topological photon transfer is achieved with the optimal shape $\alpha=1$ (sinusoidal shapes), which balances the competing effects of nonadiabatic coupling uniformity and total evolution time. This result explains why the sinusoidal shapes used in {Refs.}~\cite{JFDeng2022Science,JLYuan2021APLphotonics,ZGuan2024PRApplied,TTian2024PRB,CHWu2023PRA} enable such remarkable speed and fidelity.\medskip

\section{Competition against decoherence}
\begin{figure}[hpt]
	\centering
	\includegraphics[width=\linewidth]{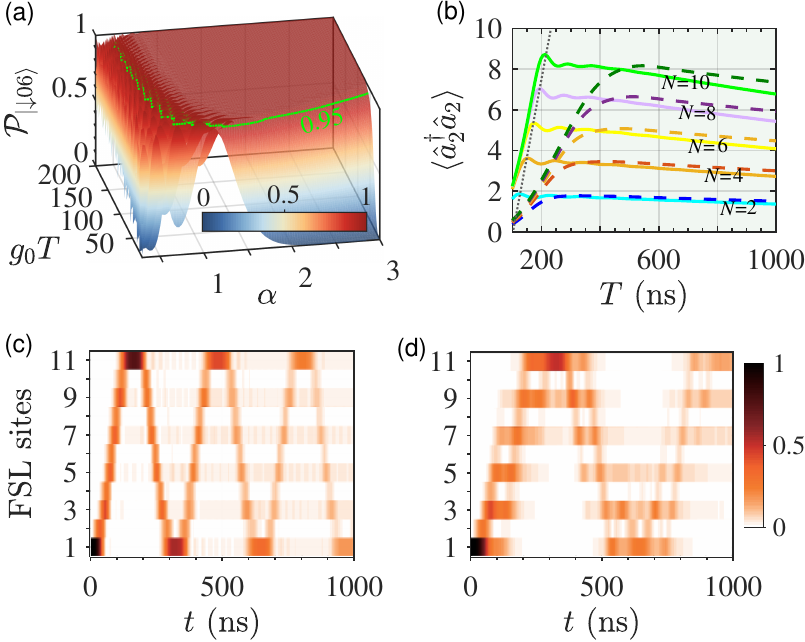}
	\caption{Numerical evaluation of topological photon transfer and decoherence effects. 
		(a) Fidelity as a function of $\alpha$ and $g_0T$ for $N=6$ (initial state $|\downarrow;6,0\rangle$). The green dashed line marks the 0.95-fidelity contour, showing that $\alpha=1$ requires the smallest $g_0T$ to reach this fidelity. 
		(b) Final average photon number $\bar n_2=\langle a_2^\dagger a_2\rangle$ versus duration $T$ for different total photon numbers $N$ (solid: $\alpha=1$, dashed: $\alpha=2$). For each curve, a peak appears due to competition between unitary transfer and decoherence. 
		(c) Time-dependent populations on the 11 FSL sites for $N=5$, $\alpha=1$, and the optimal duration $T=161$~ns. Coherent edge-to-edge pumping is observed, though even-numbered sites are also populated, indicating resonant dynamics. 
		(d) Same as (c) but with $T=322$~ns. The resonant dynamics is suppressed, and the transfer becomes more adiabatic.}\label{f3}
\end{figure}

As a crucial metric of photon transfer, the fidelity is used to assess the optimal value of $\alpha$. For the two-mode JC model with total photon number $N$, the fidelity can be defined as the population of the target Fock state $|\downarrow; 0,N\rangle$ (i.e., all photons transferred to cavity~2 from cavity~1) at the final time $T$. In Fig.~\ref{f3}a taking $N=6$ as an example, we numerically solve Schr\"odinger equation using $\hat H(t)$ to compute the topological transfer fidelity as a function of $\alpha$ and the dimensionless evolution time $g_0T$. Considering the 0.95-fidelity contour line (green dashed), it is evident that for $\alpha$ deviating from 1, achieving the same fidelity requires a larger $g_0T$, especially for $\alpha<1$. This is because the nonadiabaticity factor $Q(t)$ develops temporal fluctuations ($\sigma_Q^2>0$) when $\alpha\neq1$, leading to an increased infidelity bound. Conversely, at $\alpha=1$, the variance $\sigma_Q^2$ vanishes, and the required $g_0T$ to reach 0.95 fidelity is minimized. This result exactly corroborates our GAC theory: the optimal coupling shape that minimizes the infidelity for a given total time is the one that makes the nonadiabaticity factor as uniform as possible.\medskip

However, in the actual experiment, the transferred photon number to cavity 2 is inevitably affected by decoherence of the superconducting quantum system, especially when the transfer duration is relatively long. Key experimental parameters can be found in {Ref.}~\cite{JFDeng2022Science}, including the maximum coupling strength ($g_0/2\pi\sim9$~MHz), the coherence times of the gmon qubit ($T_1^{\text{q}} \sim 2.4$~$\mu$s and $T_2^{\ast,\text{q}} \sim 2.3$~$\mu$s), the resonator~1 ($T_1^{\text{c1}} \sim 7.9$~$\mu$s, $T_2^{\ast,\text{c1}} \sim 0.87$~$\mu$s), and the resonator~2 ($T_1^{\text{c2}} \sim 7.5$~$\mu$s, $T_2^{\ast,\text{c2}}\sim 0.90$~$\mu$s). Following standard definitions~\cite{ZLXiang2013RMP,XGu2017PR,KHe2021CPB,PYSong2026CPL,Blais2021RMP,SJMa2019LPR}, $T_1$ and $T_2^\ast$ denote the energy relaxation time and the Ramsey dephasing time of the qubit or resonators, respectively. In this situation, the evolution of the two-mode JC system is governed by the Markovian master equation
\begin{eqnarray}
\dot{{\hat\rho}}(t)&=& -i[\hat H(t),~{\hat\rho}(t)] \nonumber\\
&&+ \gamma_{\text{q,rel}} \left[ {\hat\sigma}^-{\hat\rho}(t){\hat\sigma}^+ - \frac{1}{2}\{{\hat\sigma}^+{\hat\sigma}^-,{\hat\rho}(t)\} \right] \nonumber\\
&&+ \frac{1}{2}\gamma_{\text{q},\phi} \left[ {\hat\sigma}_z{\hat\rho}(t){\hat\sigma}_z - {\hat\rho}(t) \right] \nonumber\\
&&+ \sum_{j=1}^{2} \kappa_j \left[{\hat a}_j{\hat\rho}(t) {\hat a}_j^\dagger - \frac{1}{2}\{{\hat a}_j^\dagger {\hat a}_j,{\hat\rho}(t)\} \right] \nonumber\\
&&+ \sum_{j=1}^{2} 2\gamma_{j,\phi} \left[{\hat n}_j{\hat\rho}(t) {\hat n}_j - \frac{1}{2}\{{\hat n}_j^2,{\hat\rho}(t)\} \right], \label{masterequation}
\end{eqnarray}
where ${\hat\sigma}_z = |\uparrow\rangle\langle\uparrow| - |\downarrow\rangle\langle\downarrow|$ and ${\hat n}_j = {\hat a}_j^\dagger {\hat a}_j$. $\hat\rho(t)$ is the instantaneous density operator of the system. The decoherence rates are given by: qubit energy relaxation $\gamma_{\text{q,rel}} = 1/T_1^{\text{q}}$, qubit pure dephasing $\gamma_{\text{q},\phi} = 1/T_2^{\ast,\text{q}} - 1/(2T_1^{\text{q}})$, cavity energy decay $\kappa_j = 1/T_1^{\text{c}j}$, and cavity pure dephasing $\gamma_{j,\phi} = 1/T_2^{\ast,\text{c}j} - \kappa_j/2$. \medskip

Through numerically solving the master equation for $\hat\rho(T)$, we calculate the final average photon number in cavity~2 as a function of the duration $T$, as shown in Fig.~\ref{f3}b, for different total photon numbers $N$ (initially all $N$ photons are in cavity~1). Solid and dashed lines correspond to $\alpha=1$ and $\alpha=2$, respectively. Two phenomena are clearly observed. First, the case $\alpha=1$ reaches the maximum transferred photon number significantly faster than $\alpha=2$, especially for larger $N$. Second, there is a competition between the unitary topological transfer and decoherence. The former drives the increase of $\bar n_2=\langle \hat a_2^\dagger \hat a_2 \rangle$, while the latter counteracts it. As a result, each curve exhibits a distinct peak (optimal duration) where the two effects balance.\medskip

Moreover, the optimal durations for $\alpha=1$ follow a linear relation $T_{\text{opt}}(N)=10.5N+108.4$~ns (see black dotted line in Fig.~\ref{f3}b). Remarkably, this numerical phenomenon perfectly corroborates our theoretical deduction from {Equation}~(7), demonstrating that the intrinsic $\sqrt{N}$ scaling of the nonadiabaticity factor fundamentally dictates the necessary transfer duration. This indicates that, under the considered parameters, the experimental choice $T=600$~ns for $N=5$ in {Ref.}~\cite{JFDeng2022Science} might not be optimal for maximizing the transferred photon number (the prediction gives $T_{\text{opt}}(5)=161$~ns), although it certainly guarantees better adiabaticity. For a more concrete illustration, Fig.~\ref{f1}d also shows the final average photon number in cavity~2 versus $T$ for $N=5$ in the presence of decoherence. Compared with the decoherence-free case where the optimal duration for the peak fidelity is $T=255$~ns marked by the blue pentacle, the optimal duration is reduced to $161$~ns (green triangle) due to decoherence, and the peak average photon number is lowered to $\bar n_2 = 4.4$. Relative to the experimental choice $T=600$~ns (red square, $\bar n_2 = 3.4$), the optimal duration predicted by our GAC theory reduces the transfer time by $73.2\%$ while increasing the final average photon number $\bar n_2$ by $29.4\%$. Fig.~\ref{f3}c displays the time-dependent populations on the 11 sites of the FSL for $T=161$~ns, showing coherent topological edge-to-edge pumping with a relatively higher fidelity than that reported in Fig.~2C of {Ref.}~\cite{JFDeng2022Science}. Notably, even-numbered sites are also significantly populated during the evolution, indicating resonant dynamics between the dark and bright states. This resonant dynamics is suppressed when the duration is prolonged (see Fig.~\ref{f3}d for $T=322$~ns). Therefore, the optimal duration predicted by our GAC theory results from a trade-off between speed and adiabaticity.

\section{Uniform nonadiabaticity \textit{vs.} constant gap}
Previous discussions of the fast topological transfer in the two-mode JC model have often attributed the success of the sinusoidal profiles of coupling ($\alpha=1$) to the fact that they keep the energy gap $\Delta E = \sqrt{N}g_0$ constant throughout the evolution. This argument, while plausible, misses the deeper mechanism revealed by the GAC. Indeed, the GAC shows that the infidelity bound depends not on the constancy of $\Delta E$, but on the temporal variance $\sigma_Q^2$ of the nonadiabatic coupling $Q(t)=\sqrt{N/2}|\dot\theta(t)|/\Delta E(t)$. A constant gap alone does not guarantee a small variance. It is the uniformity of $Q(t)$ that matters.\medskip

To illustrate this point, we finally consider an alternative coupling family that strictly maintains a constant gap for any $\alpha$
\begin{equation}\label{CostantGap}
g_1(t)=g_0\left|\sin^\alpha\!\left(\frac{\pi t}{2T}\right)\right|,\quad
g_2(t)=g_0\sqrt{1-\left|\sin^{2\alpha}\!\left(\frac{\pi t}{2T}\right)\right|}.
\end{equation}
Here $g_1^2+g_2^2=g_0^2$ exactly, so $\Delta E(t)=g_0$ is constant for all $\alpha$. The mixing angle becomes $\theta(t)=\arcsin[\sin^{\alpha}(\pi t/2T)]$, leading to
\begin{equation}
Q(t)=\frac{\sqrt{N}\alpha\pi}{2\sqrt2g_0T}\frac{\sin^{\alpha-1}\phi\cos\phi}{\sqrt{1-\sin^{2\alpha}\phi}}.
\end{equation}
For $\alpha=1$, the coupling strengths, equivalent to {Equation}~\eqref{coupling}, enable constant $Q(t)={\sqrt{N}\pi}/{2\sqrt2g_0T}$~($\sigma_Q=0$). For $\alpha\neq1$, however, $Q(t)$ develops a peak (if $\alpha>1$) or a dip (if $\alpha<1$) at the midpoint $\phi=\pi/4$, giving $\sigma_Q^2>0$. Therefore, even though the energy gap is perfectly constant for all $\alpha$, the infidelity bound for $\alpha\neq1$ is strictly larger than that for $\alpha=1$ due to the non-zero variance.\medskip

\begin{figure}[hpt]
	\centering
	\includegraphics[width=\linewidth]{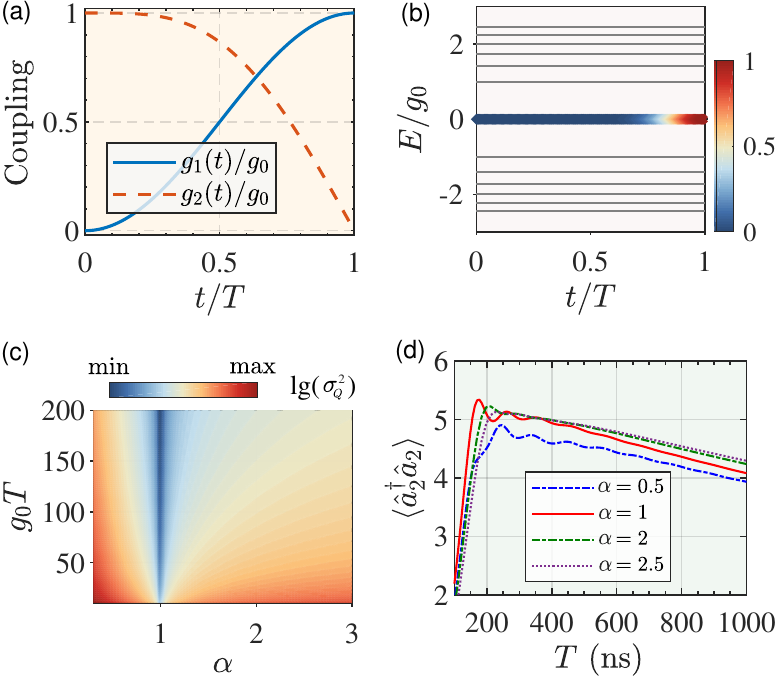}
	\caption{Constant energy gaps do not always guarantee fast transfer.
		(a) Time-dependent coupling strengths for $\alpha=2$ using the alternative coupling family in {Equation}~\eqref{CostantGap} that keeps $g_1^2+g_2^2=g_0^2$ constant for any $\alpha$. 
		(b) Corresponding eigenenergies, which are time- and $\alpha$-independent. Color on the zero-energy line indicates $|\langle \downarrow;~0,6|\psi_{\text{dark}}(t)\rangle|^2$.
		(c) Numerically computed variance $\sigma_Q^2$ of the nonadiabaticity factor as a function of $\alpha$ and $g_0T$, showing that $\sigma_Q^2=0$ only at $\alpha=1$. 
		(d) Final average photon number $\bar n_2$ versus duration $T$ for different $\alpha$ (initial state $|\downarrow;~6,0\rangle$) including decoherence. The $\alpha=1$ curve yields the shortest duration and the highest peak $\bar n_2$, confirming that uniformity of $Q(t)$ (zero variance) is the essential condition, instead of a constant gap.}\label{f4}
\end{figure}
As an illustrative example, Fig.~\ref{f4}a shows the time-dependent coupling strengths for $\alpha=2$ as defined in {Equation}~\eqref{CostantGap}. The resulting eigenenergies are constant and displayed in Fig.~\ref{f4}b. For this coupling family, varying $\alpha$ does not alter the time-independent spectrum. Fig.~\ref{f4}c numerically confirms that $\alpha=1$ yields the minimum (zero) variance of the nonadiabaticity factor $Q(t)$. Furthermore, including decoherence, we numerically compute the final average photon number in cavity~2 (initial state $|\downarrow;6,0\rangle$) versus duration $T$ for various $\alpha$, as displayed in Fig.~\ref{f4}d. Under the competition with decoherence, the predicted optimal $\alpha=1$ yields the shortest optimal duration and the highest peak average photon number $\bar n_2$ among the $\bar n_2$-$T$ curves. Therefore, the counterexample given in {Equation}~\eqref{CostantGap} conclusively demonstrates that a constant energy gap is neither necessary nor sufficient for fast, high-fidelity topological photon transfer. The essential condition is the vanishing temporal variance of the nonadiabatic coupling, that is, making $Q(t)$ as uniform as possible.

\section{Conclusion}

In summary, we have established a GAC theory to elucidate the underlying physical mechanism of fast topological photon transfer in FSLs. Contrary to the incomplete picture that a constant energy gap dictates the acceleration of topological pumping, our GAC reveals that the essential condition for minimizing infidelity is the vanishing temporal variance of the nonadiabatic coupling. \medskip

By applying this criterion to a two-mode JC model, we mathematically demonstrated that within the power-law coupling family, only the sinusoidal modulation ($\alpha = 1$) guarantees a uniform nonadiabaticity factor ($\sigma_{Q}^2 = 0$), thereby serving as the globally optimal protocol. Furthermore, by incorporating realistic decoherence parameters from recent superconducting quantum circuit experiments, we showed that the optimal transfer duration results from a direct competition between coherent pumping dynamics and system decoherence. Guided by our framework, we predicted that the topological transfer of a five-photon state can be completed in just $161$~ns, which represents a reduction of more than $73\%$ in time compared to the previous experimental implementation~\cite{JFDeng2022Science}, while simultaneously increasing the transferred photon number by over $29\%$. We also uncovered a simple linear scaling law between the optimal duration and the total photon number, providing an immediate and practical guideline for experiments. In addition, we constructed an alternative coupling family that strictly maintains a constant energy gap for any power exponent. This comparative analysis unequivocally confirmed that a constant gap alone is not sufficient for fast topological photon transfer, and non-optimal exponents still lead to larger infidelity due to non-zero variance. Therefore, the uniformity of the nonadiabatic coupling, instead of constant gaps, is the essential requirement.\medskip

This work not only offers a rigorous quantitative explanation for the surprisingly fast topological pumping observed in recent FSL experiments, but also establishes a versatile and general framework for designing fast, high-fidelity operations in topological photonics and broader quantum information processing platforms.\medskip

\section*{Acknowledgments}
The authors acknowledge supports from the National Natural Science Foundation of China (NSFC) (62571494, 12304407, 12575032, 12274376); Natural Science Foundation of Henan Province (262300421244, 262300421245, 262300422574, 232300421075); China Postdoctoral Science Foundation~(2023TQ0310, GZC20232446, 2024M762973); The Open Project of the State Key Laboratory of Metabolic Dysregulation \& Prevention and Treatment of Esophageal Cancer (2025SGAQZ-QN-05).

\bibliography{sample}
 \end{document}